\documentclass[%
 aip,
 sd,%
 amsmath,amssymb,
]{revtex4-1}

\usepackage{graphicx}
\usepackage{dcolumn}
\usepackage{bm}
\usepackage{amsthm,mathrsfs}

\newcommand{\Rm}{\mathbb{R}}
\newcommand{\Cm}{\mathbb{C}}

\newcommand{\ba}{\begin{eqnarray*}}
\newcommand{\ea}{\end{eqnarray*}}
\newcommand{\be}{\begin{equation}}
\newcommand{\ee}{\end{equation}}
\newcommand{\bea}{\begin{eqnarray}}
\newcommand{\eea}{\end{eqnarray}}
\newcommand{\va}{\varphi}
\newcommand{\pp}{\partial}
\newcommand{\vv}[1]{\boldsymbol{\mathrm{#1}}}

\usepackage{color}

\newtheorem{thm}{Theorem}[section]

\theoremstyle{remark}\newtheorem{rmk}[thm]{Remark}

\begin{document}


\title[]{The time-fractional radiative transport equation -- 
Continuous-time random walk, diffusion approximation, and 
Legendre-polynomial expansion}

\author{Manabu Machida}
\email{machida@hama-med.ac.jp}
\altaffiliation[]{Institute for Medical Photonics Research, 
Hamamatsu University School of Medicine\\
Hamamatsu, Shizuoka 431-3192, Japan}

\date{\today}

\begin{abstract}
We consider the radiative transport equation in which the time derivative 
is replaced by the Caputo derivative. Such fractional-order derivatives are 
related to anomalous transport and anomalous diffusion. In this paper we 
describe how the time-fractional radiative transport equation is obtained from 
continuous-time random walk and see how the equation is related to 
the time-fractional diffusion equation in the asymptotic limit. Then we 
solve the equation with Legendre-polynomial expansion.
\end{abstract}

\maketitle

\section{Introduction}
\label{intro}

Anomalous diffusion is often observed in nature
\cite{Metzler-etal14,Metzler-Klafter00}. For example, 
tracer particles flowing in an aquifer exhibits anomalous diffusion 
\cite{Adams-Gelhar92}. At the macroscopic scale 
after multiple scattering takes place, such anomalous diffusion is governed 
by fractional diffusion equations \cite{Metzler-Klafter00,Metzler-Klafter04,
Sokolov-Klafter-Blumen02}. 
Considering the fact that the diffusion equation appears in the asymptotic 
limit of the radiative transport equation or the linear Boltzmann equation 
\cite{Larsen-Keller74}, one can expect that at the mesoscopic scale there exist anomalous transport 
phenomena which are described by the fractional 
radiative transport equation. The use of the radiative 
transport equation was proposed for predicting the concentration of 
radionuclides in fractured rock underground \cite{Williams92,Williams93}. 
If this happens, then its fractional version must appear just like the 
fractional diffusion equation shows up when the diffusion process takes 
place in a complex structure.

Let $\alpha\in(0,1)$ and $\sigma_t,\sigma_s\in(0,\infty)$ be constants 
determined by the medium under consideration. We suppose 
$\sigma_t>\sigma_s$. Let $v>0$ be a constant speed. Let $u(x,\mu,t)$ 
($x\in\Rm$, $\mu\in[-1,1]$, $t\in[0,\infty)$) be the angular density. We 
consider the following initial-value problem for the time-fractional 
radiative transport equation. 
\be
\left\{\begin{aligned}
\pp_t^{\alpha}u(x,\mu,t)+v\mu\pp_xu(x,\mu,t)+\sigma_tu(x,\mu,t)
=\sigma_s\int_{-1}^1p(\mu,\mu')u(x,\mu',t)\,d\mu',
\\
u(x,\mu,0)=\delta(x)\delta(\mu-\mu_0),
\end{aligned}\right.
\label{intro:rte}
\ee
where $\delta(\cdot)$ is the Dirac delta function and $\pp_t^{\alpha}$ is the 
Caputo fractional derivative \cite{Caputo67}, which is defined by 
\cite{Podlubny99}
\[
\pp_t^{\alpha}u(\cdot,\cdot,t)=
\frac{1}{\Gamma(1-\alpha)}\int_0^t\frac{\pp_{t'}u(\cdot,\cdot,t')}
{(t-t')^{\alpha}}\,dt',
\qquad0<\alpha<1,
\]
with $\Gamma(\cdot)$ the Gamma function. Indeed, $u$ in (\ref{intro:rte}) 
is the fundamental solution of the time-fractional radiative transport 
equation.  We note that recently 
$\pp_t^{\alpha}$ was redefined more generally using fractional Sobolev spaces 
\cite{Gorenflo-Luchko-Yamamoto15}. Compared with the Riemann-Liouville 
derivative, the Caputo derivative is not singular at $t=0$. Thus we can have 
the same initial condition in (\ref{intro:rte}) and in the corresponding 
equation of the first derivative $\pp_t$. The phase function $p(\mu,\mu')$ 
satisfies
\[
\int_{-1}^1p(\mu,\mu')\,d\mu'=1,\qquad\forall\mu\in[-1,1].
\]

Anomalous transport phenomena 
are in the transport regime when the distance of interest is not too large 
compared to the scattering mean free path $v/\sigma_s$, and 
as is shown below, the time-fractional diffusion equation is obtained 
from (\ref{intro:rte}) in the asymptotic limit. 
The time-fractional diffusion equation has been intensively studied. 
In addition to several examples \cite{Metzler-Klafter00,Metzler-etal14}, 
we point out that the behavior of water transport in granite was successfully 
reproduced by the random walk process with a power-law distribution 
\cite{Hatano-Hatano98}. It is proposed that if there are two porosities, 
the mass transport in fractured porous aquifer should be governed by the 
diffusion equation in which both $\pp_t$ and $\pp_t^{\alpha}$ appear 
\cite{Fomin-Chugunov-Hashida11}. 
The Cauchy problem \cite{Eidelman-Kochubei04} and 
initial-boundary-value problem \cite{Luchko10,Luchko12} were considered for 
the time-fractional diffusion equation. The 
maximum principle was established \cite{Luchko09}. The technique of 
eigenfunction expansion was developed \cite{Sakamoto-Yamamoto11}. 
Numerical algorithms for the equation have been developed \cite{Lin-Xu07}. 
Moreover the standard time-fractional diffusion equation was generalized to 
equations with multiple Caputo derivatives \cite{Li-Liu-Yamamoto15,Luchko11} 
and distributed-order equations \cite{Kochubei08,Li-Luchko-Yamamoto14}. 
See the recent review by Jin and Rundell \cite{Jin-Rundell15}.

The rest of the paper is organized as follows. In \S\ref{ctrw}, we obtain the 
time-fractional radiative transport equation from continuous-time random 
walk. In \S\ref{da}, we see that the time-fractional diffusion equation 
emerges from the time-fractional radiative transport equation when absorption 
is small, propagation distance is large, and observation time is long. In 
\S\ref{fs}, we express the solution to the time-fractional radiative transport 
equation in the form of Legendre polynomial expansion. In \S\ref{num}, we 
numerically compute the solutions of the time-fractional radiative transport 
equation and of the time-fractional diffusion equation. Finally in 
\S\ref{conclusions}, concluding remarks are made. The subtraction of the 
ballistic term is considered in Appendix.

\section{Continuous-time random walk}
\label{ctrw}

We consider the continuous-time random walk whose jump probability density 
function $\va(x,t;\mu,\mu')$ ($x\in\Rm$, $t\in[0,\infty)$, 
$\mu,\mu'\in[-1,1]$) is given by
\be
\va(x,t;\mu,\mu')=
\left[\xi_s\delta(x)p(\mu,\mu')+
(1-\xi_t)\delta(x-\mu r)\delta(\mu-\mu')\right]w(t),
\label{ctrw:va}
\ee
where $\xi_t\in(0,1)$, $\xi_s\in(0,\xi_t)$, and $r>0$ are some constants. The first term represents scattering and the second term 
in the square brackets of (\ref{ctrw:va}) is responsible for transport. 
The waiting time probability density function $w(t)$ is obtained as
\[
(1-\xi_a)w(t)=\int_{-1}^1\int_{-\infty}^{\infty}\va(x,t;\mu,\mu')\,dxd\mu',
\]
where $\xi_a=\xi_t-\xi_s>0$ is the probability for absorption. The left-hand 
side of the above-mentioned equation shows the probability that the test 
particle is not absorbed in the medium and makes a jump after the time $t$.

Let $\eta(x,\mu,t)$ be the probability density function of just having arrived 
at position $x$ at time $t$ in direction $\mu$. Let $P(x,\mu,t)$ be the 
probability density function of being at 
$(x,\mu,t)\in\Rm\times[-1,1]\times[0,\infty)$. We consider the following 
continuous-time random walk process.
\[
\left\{\begin{aligned}
\eta(x,\mu,t)=\int_0^t\int_{-1}^1\int_{-\infty}^{\infty}\eta(x',\mu',t')
\va(x-x',t-t';\mu,\mu')\,dx'd\mu'dt'
+a(x,\mu)\delta(t),
\\
P(x,\mu,t)=\int_0^t\eta(x,\mu,t')\Phi(t-t')\,dt',
\end{aligned}\right.
\]
where $a(x,\mu)$ is the initial value which is a function of $x$ and $\mu$, 
$\Phi(t)$ is the cumulative probability of not having moved during $t$, 
which is given by
\[
\Phi(t)=1-\int_0^tw(t')\,dt'.
\]

By the Fourier-Laplace transform we have
\ba
(\mathcal{LF}P)(k,\mu,s)
&=&
\int_0^{\infty}e^{-st}\int_{-\infty}^{\infty}e^{-ikx}P(x,\mu,t)\,dxdt
\\
&=&
(\mathcal{LF}\eta)(k,\mu,s)(\mathcal{L}\Phi)(s),
\ea
where
\[
(\mathcal{L}\Phi)(s)=\frac{1-(\mathcal{L}w)(s)}{s}.
\]
Hence we obtain
\ba
(\mathcal{LF}\eta)(k,\mu,s)
&=&
\Biggl[\xi_s\int_{-1}^1p(\mu,\mu')(\mathcal{LF}\eta)(k,\mu',s)\,d\mu'
\\
&+&
(1-\xi_t)(\mathcal{LF}\eta)(k,\mu,s)e^{-i\mu rk}\Biggr](\mathcal{L}w)(s)
+(\mathcal{F}a)(k,\mu).
\ea
We consider small $k$ and use
\[
e^{-i\mu rk}\sim1-i\mu rk.
\]
Thus we arrive at
\ba
&&
\frac{1-(\mathcal{L}w)(s)}{(\mathcal{L}w)(s)}\left[
(\mathcal{L}P)(x,\mu,s)-\frac{1}{s}P(x,\mu,0)\right]
\\
&&=
\xi_s\int_{-1}^1p(\mu,\mu')(\mathcal{L}P)(x,\mu',s)\,d\mu'-
\left[\xi_t+(1-\xi_t)r\mu\pp_x\right](\mathcal{L}P)(x,\mu,s).
\ea

Recalling $0<\alpha<1$, we have \cite{Podlubny99,Samko93}
\[
\left(\mathcal{L}\pp_t^{\alpha}f\right)(s)
=s^{\alpha}\left(\mathcal{L}f\right)(s)-s^{\alpha-1}f(0).
\]
Let us assume that the waiting time probability density function behaves as
\[
(\mathcal{L}w)(s)\sim1-(\tau s)^{\alpha},\qquad 0<s\ll\frac{1}{\tau}.
\]
We introduce
\[
\sigma_t=\frac{\xi_t}{\tau^{\alpha}},\qquad
\sigma_s=\frac{\xi_s}{\tau^{\alpha}},\qquad
v=\frac{(1-\xi_t)r}{\tau^{\alpha}}.
\]
We asymptotically obtain
\[
\pp_t^{\alpha}P(x,\mu,t)+v\mu\pp_xP(x,\mu,t)+\sigma_tP(x,\mu,t)
=\sigma_s\int_{-1}^1p(\mu,\mu')P(x,\mu',t)\,d\mu'.
\]
This is (\ref{intro:rte}).

\begin{rmk}
In this section we implemented the effect of absorption in our 
random walk by introducing $\xi_a$. Such extension of the usual 
continuous-time random walk is done by Hornung, Berkowitz, and Barkai 
\cite{Hornung-Berkowitz-Barkai05}, and by Henry, Langlands, and Wearne 
\cite{Henry-Langlands-Wearne06}. Indeed, we arrive at the same conclusion 
by instead writing (\ref{ctrw:va}) as
\[
\va(x,t;\mu,\mu')=
\left[\xi_s\delta(x)p(\mu,\mu')+
(1-\xi_t)\delta(x-\mu r)\delta(\mu-\mu')\right]\frac{w(t)}{1-\xi_a},
\]
with the waiting time probability density function $w(t)$ introduced as
\[
w(t)=\int_{-1}^1\int_{-\infty}^{\infty}\va(x,t;\mu,\mu')\,dxd\mu'.
\]
We can then give $\eta(x,\mu,t)$ and $P(x,\mu,t)$ as
\[
\left\{\begin{aligned}
\eta(x,\mu,t)=(1-\xi_a)\int_0^t\int_{-1}^1\int_{-\infty}^{\infty}
\eta(x',\mu',t')\va(x-x',t-t';\mu,\mu')\,dx'd\mu'dt'
+a(x,\mu)\delta(t),
\\
P(x,\mu,t)=(1-\xi_a)\int_0^t\eta(x,\mu,t')\Phi(t-t')\,dt'.
\end{aligned}\right.
\]
Note that $P(x,\mu,0)=(1-\xi_a)a(x,\mu)$. Thus the relation to 
the past work \cite{Hornung-Berkowitz-Barkai05,Henry-Langlands-Wearne06} 
becomes clearer.
\end{rmk}

\section{Diffusion approximation}
\label{da}

Let us suppose that the ratio $\epsilon>0$ of the mean free path to 
the propagation distance is small. We scale $t,x$ as 
$t\rightarrow\epsilon^{2/\alpha}t$ and $x\rightarrow\epsilon x$. 
Furthermore we scale $\sigma_a\rightarrow\sigma_a/\epsilon^2$ 
assuming $\sigma_a$ is small (recall $\sigma_a=\sigma_t-\sigma_s$). 
Although the radiative transport equation (\ref{intro:rte}) has the Caputo 
derivative, we obtain the time-fractional diffusion equation by following 
the standard procedure \cite{AS09,Larsen-Keller74,Ryzhik-etal96}. In this 
section we assume that $p(\mu,\mu')=p(\mu',\mu)$. We can write the 
time-fractional radiative transport equation as
\[
\epsilon^2\pp_t^{\alpha}u(x,\mu,t)+\epsilon v\mu\pp_xu(x,\mu,t)
+\left(\epsilon^2\sigma_a+\sigma_s\right)u(x,\mu,t)
=\sigma_s\int_{-1}^1p(\mu,\mu')u(x,\mu',t)\,d\mu'.
\]
We write
\[
u(x,\mu,t)=U_{\rm DA}(x,\mu,t)+\epsilon U_{\rm DA}^{(1)}(x,\mu,t)+
\epsilon^2U_{\rm DA}^{(2)}(x,\mu,t)+\cdots.
\]
Let us collect terms of order $\epsilon^0$. We obtain
\[
\sigma_sU_{\rm DA}(x,\mu,t)=
\sigma_s\int_{-1}^1p(\mu,\mu')U_{\rm DA}(x,\mu',t)\,d\mu'.
\]
The above equation implies that $U_{\rm DA}$ is independent of $\mu$; 
hereafter we write $U_{\rm DA}(x,\mu,t)=U_{\rm DA}(x,t)$. 
The terms of order $\epsilon^1$ yields
\[
v\mu\pp_xU_{\rm DA}(x,t)+\sigma_sU_{\rm DA}^{(1)}(x,\mu,t)=
\sigma_s\int_{-1}^1p(\mu,\mu')U_{\rm DA}^{(1)}(x,\mu',t)\,d\mu'.
\]
We obtain
\[
U_{\rm DA}^{(1)}(x,\mu,t)=
-\frac{v}{(1-\mathrm{g})\sigma_s}\mu\pp_xU_{\rm DA}(x,t),
\]
where $\mathrm{g}\in(-1,1)$ satisfies
\[
\mu\mathrm{g}=\int_{-1}^1\mu'p(\mu,\mu')\,d\mu'.
\]
By collecting terms of order $\epsilon^2$ we have
\ba
&&
\pp_t^{\alpha}U_{\rm DA}(x,t)+\mu\pp_xU_{\rm DA}^{(1)}(x,\mu,t)+
\sigma_aU_{\rm DA}(x,t)+\sigma_sU_{\rm DA}^{(2)}(x,\mu,t)
\\
&&=
\sigma_s\int_{-1}^1p(\mu,\mu')U_{\rm DA}^{(2)}(x,\mu',t)\,d\mu'.
\ea
If we integrate the above equation over $\mu$, we obtain
\be
\pp_t^{\alpha}U_{\rm DA}(x,t)-D_0\pp_x^2U_{\rm DA}(x,t)+
\sigma_aU_{\rm DA}(x,t)=0,
\label{da:diffusioneq}
\ee
where
\be
D_0=\frac{v}{3(1-\mathrm{g})\sigma_s}.
\label{da:D0}
\ee
Thus the time-fractional diffusion equation is obtained in the asymptotic 
limit of (\ref{intro:rte}).

One remark needs to be made. We have the second derivative for the spatial 
variable $x$ in (\ref{da:diffusioneq}). In a similar setting, it is known that the space-fractional diffusion 
equation is obtained if the phase function decays with power-law 
as a function of the speed of propagating particles 
\cite{Mellet10,Mellet-Mischler-Mouhot11}.

\section{Legendre-polynomial expansion}
\label{fs}

Let us suppose $p(\mu,\mu')$ is given by
\[
p(\mu,\mu')=\frac{1}{2}\sum_{l=0}^L\beta_lP_l(\mu)P_l(\mu'),
\]
where $L\ge0$, and $\beta_l$ ($l=0,1,\dots,L$) are positive constants such as
$\beta_0=1$, $\beta_l<2l+1$ for $l\ge1$. Here, $P_l(\mu)$ are the Legendre 
polynomials recursively given by
\[
(l+1)P_{l+1}(\mu)=(2l+1)\mu P_l(\mu)-lP_{l-1}(\mu),\qquad
P_1(\mu)=\mu,\qquad P_0(\mu)=1,\qquad\mu\in[-1,1].
\]
In the time-independent case, an analytical solution of the 
space-fractional radiative transport equation was found 
\cite{Kadem-Luchko-Baleanu10}. 
In this section we solve (\ref{intro:rte}). 
Let us expand $u$ with Legendre polynomials.
\be
(\mathcal{F}u)(k,\mu,t)
=\sum_{l=0}^{\infty}\sqrt{2l+1}c_l(k,t;\mu_0)P_l(\mu).
\label{fs:expand}
\ee
We perform the Fourier transform in (\ref{intro:rte}) and substitute 
(\ref{fs:expand}). We have
\ba
&&
\left(\pp_t^{\alpha}+ivk\mu+\sigma_t\right)
\sum_{l=0}^{\infty}\sqrt{2l+1}c_l(k,t;\mu_0)P_l(\mu)
\\
&&=
\sigma_s\sum_{l=0}^{\infty}
\sqrt{2l+1}c_l(k,t;\mu_0)\frac{\beta_l}{2l+1}P_l(\mu)\Theta(L-l).
\ea
Let us introduce
\[
h_l=2l+1-\frac{\sigma_s}{\sigma_t}\beta_l\Theta(L-l).
\]
Let $N$ ($\ge L$) be an integer. We take projections with $P_l(\mu)$ 
($l=0,1,\dots,N$) and obtain
\[
\frac{ivkl}{\sqrt{4l^2-1}}c_{l-1}+\frac{ivk(l+1)}{\sqrt{4(l+1)^2-1}}c_{l+1}+
\pp_t^{\alpha}c_l+\frac{\sigma_th_l}{2l+1}c_l=0,
\]
where we used the recurrence relations and orthogonality relations 
of Legendre polynomials,
\be
\mu P_l(\mu)=\frac{l+1}{2l+1}P_{l+1}(\mu)+\frac{l}{2l+1}P_{l-1}(\mu),
\label{threeterm}
\ee
and
\[
\int_{-1}^1P_l(\mu)P_{l'}(\mu)\,d\mu=\frac{2}{2l+1}\delta_{ll'}.
\]
The above equation is expressed as
\[
A(k)\vv{c}(k,t;\mu_0)+\pp_t^{\alpha}\vv{c}(k,t;\mu_0)=0,
\]
where $A(k)$ is an $(N+1)\times(N+1)$ matrix and $\vv{c}(k,t;\mu_0)$ is 
an $N+1$ dimensional vector defined by
\bea
\{A(k)\}_{ll'}=
\frac{ivkl}{\sqrt{4l^2-1}}\delta_{l-1,l'}+
\frac{\sigma_th_l}{2l+1}\delta_{l,l'}+
\frac{ivk(l+1)}{\sqrt{4(l+1)^2-1}}\delta_{l+1,l'},
\label{fs:Amat}
\\
\{\vv{c}(k,t;\mu_0)\}_l=
c_l(k,t;\mu_0).
\label{fs:coeffvec}
\eea
When the Legendre polynomial expansion is used, tridiagonal matrices such as 
$A(k)$ appear due to the three-term recurrence relation (\ref{threeterm}) 
\cite{Garcia-Siewert89,Gershenson99,Liemert-Kienle12,Panasyuk-etal06}. 
By taking the Laplace transform we have
\[
(\mathcal{L}\vv{c})(k,s;\mu_0)=
\left(A(k)+s^{\alpha}\right)^{-1}s^{\alpha-1}\vv{c}(k,0;\mu_0),
\]
where we used
\[
(\mathcal{L}\pp_t^{\alpha}\vv{c})(k,s;\mu_0)=
s^{\alpha}(\mathcal{L}\vv{c})(k,s;\mu_0)-s^{\alpha-1}\vv{c}(k,0;\mu_0),\qquad
0<\alpha\le1.
\]
Let us recall that the Mittag-Leffler function is given by 
\cite{Podlubny99}
\[
E_{\alpha}(z):=\sum_{n=0}^{\infty}\frac{z^n}{\Gamma(\alpha n+1)},\qquad
z,\alpha\in\Cm,\quad\Re{\alpha}>0,
\]
and the Laplace transform is obtained as
\[
\mathcal{L}\left\{E_{\alpha}(zt^{\alpha});\,s\right\}
=\frac{s^{\alpha-1}}{s^{\alpha}-z},\qquad
z,s,\alpha\in\Cm,\quad\Re{s},\Re{\alpha}>0,\quad
\left|\frac{z}{s^{\alpha}}\right|<1.
\]
Thus we find
\[
\vv{c}(k,t;\mu_0)=E_{\alpha}\left(-A(k)t^{\alpha}\right)\vv{c}(k,0;\mu_0).
\]

Since 
$\delta(\mu-\mu_0)=\sum_{l=0}^{\infty}\frac{2l+1}{2}P_l(\mu)P_l(\mu_0)$, 
we obtain
\[
\left\{\vv{c}(k,0;\mu_0)\right\}_l=\frac{\sqrt{2l+1}}{2}P_l(\mu_0).
\]
Let $\lambda_n(k)$ and $\vv{v}_n(k)$ be the $n$th eigenvalue and eigenvector 
of the matrix $A(k)$. We can write $A(k)$ as
\[
A(k)=Q(k)D(k)Q(k)^{-1},
\]
where
\[
Q(k)=\left(\vv{v}_0(k)\;\vv{v}_1(k)\;\cdots\;\vv{v}_N(k)\right),\qquad
D(k)=
\mathop{\mathrm{diag}}(\lambda_0(k),\lambda_1(k),\dots,\lambda_N(k)).
\]
We have
\[
\{A(k)\}_{ij}=\left\{Q(k)D(k)Q(k)^{-1}\right\}_{ij}
=\sum_{n=0}^N\lambda_n(k)v_n^{(i)}(k)v_n^{(j)*}(k),
\]
where $v_n^{(i)}(k)$ is the $i$th component of $\vv{v}_n(k)$. Therefore we 
can write
\[
\left\{\vv{c}(k,t;\mu_0)\right\}_l
=\sum_{j=0}^N\frac{\sqrt{2j+1}}{2}P_j(\mu_0)\sum_{n=0}^N
v_n^{(l)}(k)v_n^{(j)*}(k)E_{\alpha}\left(-\lambda_n(k)t^{\alpha}\right).
\]
Noting (\ref{fs:coeffvec}), Eq.~(\ref{fs:expand}) yields
\bea
u(x,\mu,t)
&\approx&
u(x,\mu,t;N)
\nonumber \\
&:=&
\frac{1}{2\pi}\int_{-\infty}^{\infty}e^{ikx}\sum_{l=0}^N\sqrt{2l+1}
c_l(k,t;\mu_0)P_l(\mu)\,dk.
\label{fs:result}
\eea
Since $k$ appears always as $ik$, we see
\[
c_l(-k,t;\mu_0)=c_l(k,t;\mu_0)^*.
\]
We obtain
\bea
u(x,\mu,t;N)
&=&
\sum_{l=0}^N\frac{\sqrt{2l+1}}{\pi}P_l(\mu)
\nonumber \\
&\times&
\int_0^{\infty}\left[\cos{(kx)}\Re{c_l(k,t;\mu_0)}-
\sin{(kx)}\Im{c_l(k,t;\mu_0)}\right]\,dk.
\label{fs:result2}
\eea

\begin{rmk}
Although in this section we directly calculated $u$ in (\ref{fs:result2}), 
indeed, it is possible to directly relate $u(x,\mu,t)$ to $u_1(x,\mu,t)$ 
which is the solution of (\ref{intro:rte}) with $\alpha=1$. 
Let $f_{\alpha}(t)$ be a function such that
\[
(\mathcal{L}f_{\alpha})(s)=e^{-s^{\alpha}}.
\]
For example, we have
\[
f_{1/2}(t)=\frac{t^{-3/2}}{2\sqrt{\pi}}e^{-1/(4t)}.
\]
If we introduce
\[
\va(\tau,t)=\frac{t}{\alpha\tau^{1+1/\alpha}}f_{\alpha}\left(
\frac{t}{\tau^{1/\alpha}}\right),
\]
we have
\[
(\mathcal{L}\va)(\tau,s)=s^{\alpha-1}e^{-\tau s^{\alpha}}.
\]
Let us consider the Laplace transform of $u$ with respect to $s$ and $u_1$ 
with respect to $s^{\alpha}$. Assuming $u(x,\mu,0)=u_1(x,\mu,0)$, we obtain
\[
\left\{\begin{aligned}
s^{\alpha}(\mathcal{L}u)(x,\mu,s)-s^{\alpha-1}u(x,\mu,0)+
v\mu\pp_x(\mathcal{L}u)(x,\mu,s)
+\sigma_t(\mathcal{L}u)(x,\mu,x)
\\
\qquad=
\sigma_s\int_{-1}^1p(\mu,\mu')(\mathcal{L}u)(x,\mu',s)\,d\mu',
\\
s^{\alpha}(\mathcal{L}u_1)(x,\mu,s^{\alpha})-u_1(x,\mu,0)+
v\mu\pp_x(\mathcal{L}u_1)(x,\mu,s^{\alpha})+
\sigma_t(\mathcal{L}u_1)(x,\mu,s^{\alpha})
\\
\qquad=
\sigma_s\int_{-1}^1p(\mu,\mu')(\mathcal{L}u_1)(x,\mu',s^{\alpha})\,d\mu'.
\end{aligned}\right.
\]
The above equations imply
\[
(\mathcal{L}u)(x,\mu,s)=s^{\alpha-1}(\mathcal{L}u_1)(x,\mu,s^{\alpha})
=\int_0^{\infty}u_1(x,\mu,t)s^{\alpha-1}e^{-\tau s^{\alpha}}\,d\tau.
\]
Therefore $u$ and $u_1$ are related as
\[
u(x,\mu,t)=\int_0^{\infty}u_1(x,\mu,\tau)\va(\tau,t)\,d\tau.
\]
This means that we can obtain $u$ by integrating $u_1$, which is the solution 
of the first-order equation. The solution $u$ is subordinated to the solution 
$u_1$ \cite{Langlands-Henry-Wearne09}.
\end{rmk}

\section{Numerical calculation}
\label{num}

The energy density $U(x,t)$ is introduced as
\[
U(x,t)=\int_{-1}^1u(x,\mu,t)\,d\mu.
\]
Each $N$ gives an approximated value of $U(x,t)$ as
\[
U(x,t)\approx U(x,t;N),
\]
where
\[
U(x,t;N)=\int_{-1}^1u(x,\mu,t;N)\,d\mu.
\]
We note that $U(x,t)=U(x,t;\infty)$. 
Let us calculate $U(x,t;N)$ for the initial condition
\[
U(x,0;N)=\delta(x).
\]
From (\ref{fs:result2}) we obtain
\ba
&&
U(x,t;N)
=
\int_{-1}^1\int_{-1}^1u(x,\mu,t;N)\,d\mu d\mu_0
\\
&&=
\frac{1}{\pi}\int_{-\infty}^{\infty}e^{ikx}\sum_{n=0}^N
\left|v_n^{(0)}(k)\right|^2
E_{\alpha}\left(-\lambda_n(k)t^{\alpha}\right)\,dk
\\
&&=
\frac{2}{\pi}\sum_{n=0}^N\int_0^{\infty}\left|v_n^{(0)}(k)\right|^2
\\
&&\times
\Bigl(\cos{(kx)}\Re{E_{\alpha}\left(-\lambda_n(k)t^{\alpha}\right)}-
\sin{(kx)}\Im{E_{\alpha}\left(-\lambda_n(k)t^{\alpha}\right)}\Bigr)\,dk.
\ea

In this section we set
\[
v=1,\qquad\sigma_a=0,\qquad L=N=1,
\]
and
\[
\sigma_s=10,\qquad\mathrm{g}=\frac{\beta_1}{3}=0.9.
\]
The matrix $A(k)$ in (\ref{fs:Amat}) is given by
\[
A(k)=\frac{1}{\sqrt{3}}\left(\begin{array}{cc}
0 & ik \\ ik & 2k_c
\end{array}\right).
\]
where we introduced
\[
k_c:=\frac{\sqrt{3}}{2}\sigma_s(1-\mathrm{g}).
\]
Its eigenvalues and eigenvectors are obtained as
\[
\lambda(k)=\frac{k_c}{\sqrt{3}}
\left(1\pm\sqrt{1-\left(\frac{k}{k_c}\right)^2}\right),
\]
and
\[
\vv{v}(k)=\frac{1}{\sqrt{\mathcal{N}}}\left(\begin{array}{c}
\frac{ik}{\sqrt{3}}\\\lambda(k)\end{array}\right),\qquad
\mathcal{N}=\left\{\begin{aligned}
\frac{2k_c^2}{3}\left(1\pm\sqrt{1-\left(\frac{k}{k_c}\right)^2}\right),
&\quad |k|\le k_c,
\\
\frac{2}{3}k^2,
&\quad |k|>k_c.
\end{aligned}\right.
\]
Thus we have
\[
\left|v^{(0)}(k)\right|^2
=\left\{\begin{aligned}
\frac{1}{2}\left(1\mp\sqrt{1-\left(\frac{k}{k_c}\right)^2}\right),
&\quad |k|\le k_c,
\\
\frac{1}{2},
&\quad |k|>k_c.
\end{aligned}\right.
\]
The energy density is written as
\bea
U(x,t;1)
&=&
\frac{1}{\pi}\int_0^{k_c}\cos{(kx)}
\nonumber \\
&\times&
\Biggl[\left(1-\sqrt{1-\left(\frac{k}{k_c}\right)^2}\right)
E_{\alpha}\left(-\frac{k_c+\sqrt{k_c^2-k^2}}{\sqrt{3}}t^{\alpha}\right)
\nonumber \\
&+&
\left(1+\sqrt{1-\left(\frac{k}{k_c}\right)^2}\right)
E_{\alpha}\left(-\frac{k_c-\sqrt{k_c^2-k^2}}{\sqrt{3}}t^{\alpha}\right)
\Biggr]\,dk
\nonumber \\
&+&
\frac{2}{\pi}\int_{k_c}^{\infty}\cos{(kx)}
\Re{E_{\alpha}\left(-\frac{k_c-i\sqrt{k^2-k_c^2}}{\sqrt{3}}t^{\alpha}\right)}
\,dk.
\label{num:U1}
\eea

In the diffusion approximation the energy density is given as follows. 
If the initial condition is given by
\[
U_{\rm DA}(x,0)=\delta(x),
\]
we have \cite{Mainardi96,Mainardi-Luchko-Pagnini01}
\bea
U_{\rm DA}(x,t)
&=&
\frac{1}{\pi}\int_0^{\infty}\cos{(kx)}E_{\alpha}(-D_0k^2t^{\alpha})\,dk
\label{num:UDA}
\\
&=&
\frac{1}{\sqrt{D_0}}t^{-\frac{\alpha}{2}}
M_{\alpha/2}\left(\frac{|x|}{\sqrt{D_0}t^{\alpha/2}}\right),
\nonumber
\eea
where $M_{\alpha}(z)$ is the $M$-Wright function defined by
\[
M_{\alpha}(z):=
\sum_{n=0}^{\infty}\frac{(-1)^nz^n}{n!\Gamma(-\alpha(n+1)+1)}.
\]

\begin{figure}[ht]
\begin{center}
\includegraphics[width=0.3\textwidth]{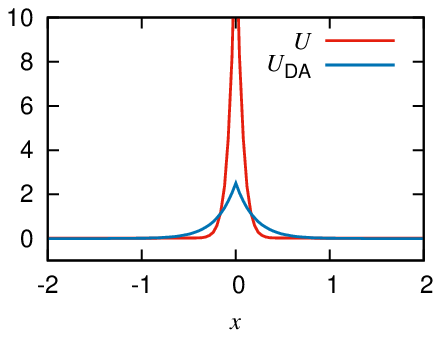}
\includegraphics[width=0.3\textwidth]{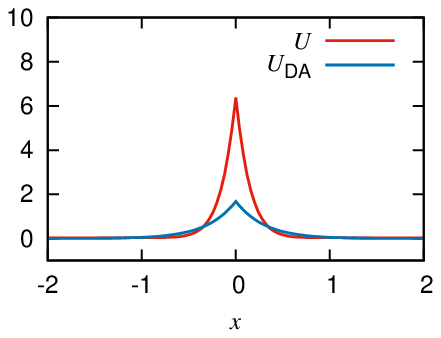}
\includegraphics[width=0.3\textwidth]{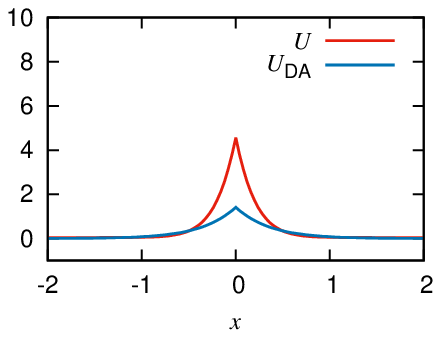}
\vspace{10mm}
\end{center}
\caption{
Comparison of $U(x,t)$ and $U_{\rm DA}(x,t)$ as a function of $x$, 
from the left, for $t=0.0001$, $0.0025$, and $0.01$, respectively 
when $\alpha=0.25$.
}
\label{fig1}
\end{figure}

\begin{figure}[ht]
\begin{center}
\includegraphics[width=0.3\textwidth]{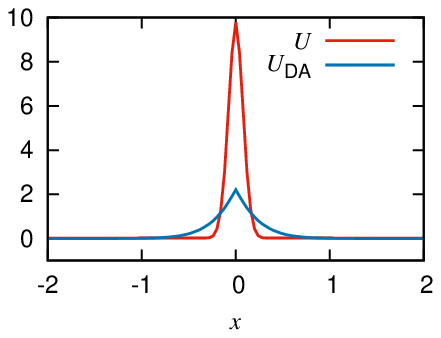}
\includegraphics[width=0.3\textwidth]{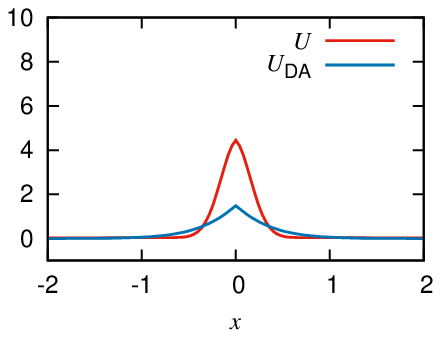}
\includegraphics[width=0.3\textwidth]{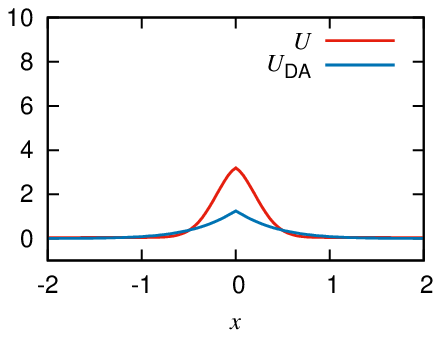}
\vspace{10mm}
\end{center}
\caption{
Comparison of $U(x,t)$ and $U_{\rm DA}(x,t)$ as a function of $x$, 
from the left, for $t=0.01$, $0.05$, and $0.1$, respectively 
when $\alpha=0.5$.
}
\label{fig2}
\end{figure}

\begin{figure}[ht]
\begin{center}
\includegraphics[width=0.3\textwidth]{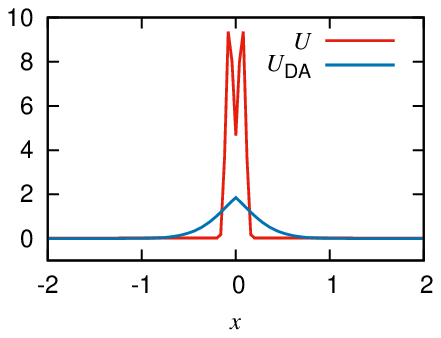}
\includegraphics[width=0.3\textwidth]{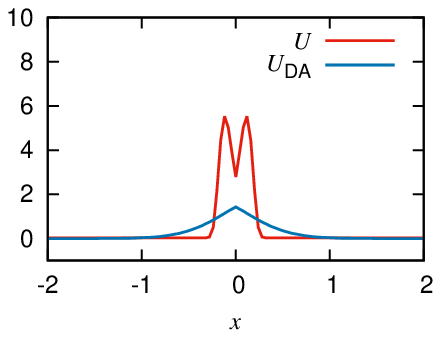}
\includegraphics[width=0.3\textwidth]{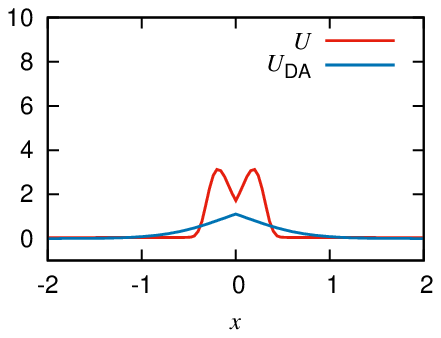}
\vspace{10mm}
\end{center}
\caption{
Comparison of $U(x,t)$ and $U_{\rm DA}(x,t)$ as a function of $x$, 
from the left, for $t=0.05$, $0.1$, and $0.2$, respectively 
when $\alpha=0.75$.
}
\label{fig3}
\end{figure}

Equations (\ref{num:U1}) and (\ref{num:UDA}) are implemented in Fortran. 
The numerical implementation of the Mittag-Lifter function relies on the 
algorithm by Gorenflo, Loutchko, and Luchko \cite{Gorenflo-Loutchko-Luchko02}. 
Although we saw in \S\ref{da} that $U(x,t)$ asymptotically becomes 
$U_{\rm DA}(x,t)$, they are different in general. 
In Figs.~\ref{fig1} through \ref{fig3}, we plot $U(x,t;1)$ and 
$U_{\rm DA}(x,t)$ for $\alpha=0.25$, $0.5$, and $0.75$, respectively. 
For all the cases, we see that $U(x,t;1)$ stays near the source at $x=0$ 
for a relatively long time whereas $U_{\rm DA}(x,t)$ broadens quickly. 
When $\alpha=0.75$ we can see that $U(x,t;1)$ has two peaks. Such a 
double-peak structure shows up for $\alpha>1$ in the case of the fractional diffusion equation 
\cite{Mainardi-Luchko-Pagnini01}. This behavior can be understood from the 
relation \cite{Erdelyi-etal55}
\[
E_{\alpha}(z)+E_{\alpha}(-z)=2E_{2\alpha}(z^2),\qquad z\in\Cm.
\]
For sufficiently large $k$, which corresponds to small $x$, we asymptotically 
have \cite{Podlubny99}
\[
E_{\alpha}\left(-\frac{k_c-i\sqrt{k^2-k_c^2}}{\sqrt{3}}t^{\alpha}\right)
\sim
\frac{1}{\alpha}\exp\left[
\left(-\frac{k_c-i\sqrt{k^2-k_c^2}}{\sqrt{3}}t^{\alpha}\right)^{1/\alpha}
\right].
\]
Hence in (\ref{num:U1}) we have
\ba
\Re{E_{\alpha}\left(-\frac{k_c-i\sqrt{k^2-k_c^2}}{\sqrt{3}}t^{\alpha}\right)}
&\sim&
\frac{1}{2\alpha}
\exp\left[\left(i\frac{k}{\sqrt{3}}t^{\alpha}\right)^{1/\alpha}\right]
+\frac{1}{2\alpha}
\exp\left[\left(-i\frac{k}{\sqrt{3}}t^{\alpha}\right)^{1/\alpha}\right]
\\
&\sim&
\frac{1}{2}E_{\alpha}\left(i\frac{k}{\sqrt{3}}t^{\alpha}\right)
+\frac{1}{2}E_{\alpha}\left(-i\frac{k}{\sqrt{3}}t^{\alpha}\right)
\\
&=&
E_{2\alpha}\left(-\frac{1}{3}k^2t^{2\alpha}\right).
\ea
The above calculation implies that the double-peak behavior for the 
fractional diffusion equation with $\alpha>1$ can be seen for the fractional 
radiative transport equation with $\alpha>1/2$.

\section{Concluding remarks}
\label{conclusions}

One of the purposes of the present paper is to see the connection between 
the time-fractional radiative transport equation and the time-fractional 
diffusion equation. Roughly speaking, the time-fractional radiative transport 
equation of $\pp_t^{\alpha}$ behaves as the time-fractional diffusion 
equation of $\pp_t^{\alpha}$ for large $x$ and behaves as the 
time-fractional diffusion equation of $\pp_t^{2\alpha}$ near $x=0$ as is 
investigated in \S\ref{da} and \S\ref{num}.

When $u(x,\mu,t)$ in (\ref{intro:rte}) is expressed in the form of the 
collision expansion, the ballistic term is singular. If $u(x,\mu,t)$ itself 
is numerically computed, it is desirable to subtract the ballistic term. 
In a straightforward manner, we can extend the calculation in 
\S\ref{fs}. This calculation is summarized in Appendix.


\appendix

\section{Subtraction of the ballistic term}
\label{bt}

Let us split $u(x,\mu,t)$ in (\ref{intro:rte}) into the ballistic and 
scattered parts as
\[
u(x,\mu,t)=u_b(x,\mu,t)+u_s(x,\mu,t),
\]
where $u_b(x,\mu,t)$ and $u_s(x,\mu,t)$ respectively satisfy
\[
\left\{\begin{aligned}
\pp_t^{\alpha}u_b(x,\mu,t)+\mu\pp_xu_b(x,\mu,t)+\sigma_tu_b(x,\mu,t)
=0,
\\
u_b(x,\mu,0)=\delta(x)\delta(\mu-\mu_0),
\end{aligned}\right.
\]
and
\[
\left\{\begin{aligned}
\pp_t^{\alpha}u_s(x,\mu,t)+\mu\pp_xu_s(x,\mu,t)+\sigma_tu_s(x,\mu,t)
=\sigma_s\int_{-1}^1p(\mu,\mu')u_s(x,\mu',t)\,d\mu'
\\
+S(x,\mu,t),
\\
u_s(x,\mu,0)=0.
\end{aligned}\right.
\]
Here the source term for $u_s(x,\mu,t)$ is given by
\[
S(x,\mu,t;\mu_0)=
\sigma_s\int_{-1}^1p(\mu,\mu')u_b(x,\mu',t)\,d\mu'.
\]
Noting that
\[
(\mathcal{LF}u_b)(k,\mu,s)=
\frac{s^{\alpha-1}}{s^{\alpha}+ik\mu+\sigma_t}\delta(\mu-\mu_0),
\]
we obtain
\[
u_b(x,\mu,t)
=\frac{1}{2\pi}\delta(\mu-\mu_0)\int_{-\infty}^{\infty}e^{ikx}
E_{\alpha}\left[-(ik\mu_0+\sigma_t)t^{\alpha}\right]\,dk,
\]
and
\[
(\mathcal{LF}S)(k,\mu,s;\mu_0)
=\sigma_sp(\mu,\mu_0)\frac{s^{\alpha-1}}{s^{\alpha}+ik\mu_0+\sigma_t}.
\]

Let us expand $u_s$ with Legendre polynomials.
\be
(\mathcal{F}u_s)(k,\mu,t)=\sum_{l=0}^{\infty}\sqrt{2l+1}c_l(k,t;\mu_0)P_l(\mu).
\label{A:expand}
\ee
For $0\le l\le N$ we obtain
\[
A(k)\vv{c}(k,t;\mu_0)+\pp_t^{\alpha}\vv{c}(k,t;\mu_0)=\vv{w}(k,t;\mu_0),
\]
where $\vv{w}(k,t;\mu_0)$ is an $N+1$ dimensional vector defined by
\[
\{\vv{w}(k,t;\mu_0)\}_l=
\frac{\sqrt{2l+1}}{2}\int_{-1}^1P_l(\mu)(\mathcal{F}S)(k,\mu,t;\mu_0)\,d\mu.
\]
By taking the Laplace transform we have
\[
(\mathcal{L}\vv{c})(k,s;\mu_0)=
\left(A(k)+s^{\alpha}\right)^{-1}\left[s^{\alpha-1}\vv{c}(k,0;\mu_0)
+(\mathcal{L}\vv{w})(k,s;\mu_0)\right].
\]
Let us express the Laplace transform of $\vv{w}(k,t;\mu_0)$ as
\[
(\mathcal{L}\vv{w})(k,s;\mu_0)=
\frac{s^{\alpha-1}}{s^{\alpha}+ik\mu_0+\sigma_t}\vv{b}(\mu_0),
\]
where
\[
\{\vv{b}(\mu_0)\}_l=\frac{\sigma_s\beta_l}{2\sqrt{2l+1}}\Theta(L-l)P_l(\mu_0).
\]
Using the relation
\ba
\left(A(k)-s^{\alpha}\right)^{-1}(\mathcal{L}\vv{w})(k,s;\mu_0)
\\
=
\left(A(k)+ik\mu_0+\sigma_t\right)^{-1}
\left(\frac{s^{\alpha-1}}{s^{\alpha}+ik\mu_0+\sigma_t}
-\frac{s^{\alpha-1}}{s^{\alpha}-A(k)}\right)\vv{b}(\mu_0),
\ea
we find
\ba
\vv{c}(k,t;\mu_0)
=
E_{\alpha}(-A(k)t^{\alpha})\vv{c}(k,0;\mu_0)
\\
+
\left(A(k)+ik\mu_0+\sigma_t\right)^{-1}\left[
E_{\alpha}\left(-(ik\mu_0+\sigma_t)t^{\alpha}\right)
-E_{\alpha}\left(A(k)t^{\alpha}\right)\right]\vv{b}(\mu_0).
\ea


\begin{thebibliography}{99}

\bibitem{Adams-Gelhar92}
Adams, E. E. and Gelhar, L. W.,
``Field study of dispersion in a heterogeneous aquifer 2. Spatial moments analysis,''
{\it Water Res. Res.} {\bf 28}, 3293--3307 (1992).

\bibitem{AS09}
Arridge, S. R. and Schotland, J. C.,
``Optical tomography: forward and inverse problems,''
{\it Inverse Problems} {\bf 25}, 123010 (2009).

\bibitem{Caputo67}
Caputo, M.,
``Linear model of dissipation whose $Q$ is almost frequency independent-II,''
{\it Geophys. J. R. Astr. Soc.} {\bf 13}, 529--539 (1967).

\bibitem{Eidelman-Kochubei04}
Eidelman, S. D. and Kochubei, A. N.,
``Cauchy problem for fractional diffusion equations,''
{\it J. Diff. Eq.} {\bf 199}, 211--255 (2004).

\bibitem{Erdelyi-etal55}
Erd\'{e}lyi, A., Magnus, W., Oberhettinger, F., and Tricomi, F. G.,
{\it Higher Transcendental Functions} Vol.~3
(McGraw-Hill, 1955).

\bibitem{Fomin-Chugunov-Hashida11}
Fomin, S. A., Chugunov, V. A., and Hashida, T.,
``Non-Fickian mass transport in fractured porous media,''
{\it Adv. Water Resour.} {\bf 34}, 205--214 (2011).

\bibitem{Garcia-Siewert89}
Garcia, R. D. M. and Siewert C. E.,
``On discrete spectrum calculations in radiative transfer,''
{\it J. Quant. Spec. Rad. Trans.} {\bf 42}, 385--394 (1989).

\bibitem{Gershenson99}
Gershenson, M.,
``Time-dependent equation for the intensity in the diffusion limit using a higher-order angular expansion,''
{\it Phys. Rev. E} {\bf 59}, 7178--7184 (1999).

\bibitem{Gorenflo-Loutchko-Luchko02}
Gorenflo, R., Loutchko, J., and Luchko, Y.,
``Computation of the Mittag-Leffler function $E_{\alpha,\beta}(z)$ and its derivative,''
{\it Fract. Calc. Appl. Anal.} {\bf 5}, 491--518 (2002).

\bibitem{Gorenflo-Luchko-Yamamoto15}
Gorenflo, R., Luchko, Y., and Yamamoto, M.,
``Time-fractional diffusion equation in the fractional Sobolev spaces,''
{\it Fract. Calc. Appl. Anal.} {\bf 18}, 799--820 (2015).

\bibitem{Hatano-Hatano98}
Hatano, Y. and Hatano, N.,
``Dispersive transport of ions in column experiments: An explanation of long-tailed profiles,''
{\it Water Resour. Res.} {\bf 34}, 1027--1033 (1998).

\bibitem{Henry-Langlands-Wearne06}
Henry, B. I., Langlands, T. A. M., and Wearne, S. L.,
``Anomalous diffusion with linear reaction dynamics: From continuous time random walks to fractional reaction-diffusion equations,''
{\it Phys. Rev. E} {\bf 74}, 031116 (2006)

\bibitem{Hornung-Berkowitz-Barkai05}
Hornung, G., Berkowitz, B., and Barkai, N.,
``Morphogen gradient formation in a complex environment: An anomalous diffusion model,''
{\it Phys. Rev. E} {\bf 72}, 041916 (2005).

\bibitem{Jin-Rundell15}
Jin, B. and Rundell, W.,
``A tutorial on inverse problems for anomalous diffusion processes,''
{\it Inverse Problems} {\bf 31}, 035003 (2015).

\bibitem{Kadem-Luchko-Baleanu10}
Kadem, A., Luchko, Y., and Baleanu, D.,
``Spectral method for solution of the fractional transport equation,''
{\it Rep. Math. Phys.} {\bf 66}, 103--115 (2010).

\bibitem{Kochubei08}
Kochubei, A. N.,
``Distributed order calculus and equations of ultraslow diffusion,''
{\it J. Math. Anal. Appl.} {\bf 340}, 252--281 (2008).

\bibitem{Langlands-Henry-Wearne09}
Langlands, T. A. M., Henry, B. I., and Wearne, S. L.,
``Fractional cable equation models for anomalous electrodiffusion in nerve cells: infinite domain solutions,''
{\it J. Math. Biol.} {\bf 59}, 761--808 (2009).

\bibitem{Larsen-Keller74}
Larsen, E. W. and Keller, J. B.,
``Asymptotic solution of neutron transport problems for small mean free paths,''
{\it J. Math. Phys.} {\bf 15}, 75--81 (1974).

\bibitem{Li-Liu-Yamamoto15}
Li, Z., Liu, Y., and Yamamoto, M.,
``Initial-boundary value problems for multi-term time-fractional diffusion equations with positive constant coefficients,''
{\it Appl. Math. Comp.} {\bf 257}, 381--397 (2015).

\bibitem{Li-Luchko-Yamamoto14}
Li, Z., Luchko, Y., and Yamamoto, M.,
``Asymptotic estimates of solutions to initial-boundary-value problems for distributed order time-fractional diffusion equations,''
{\it Fract. Cal. Appl. Anal.} {\bf 17}, 1114--1136 (2014).

\bibitem{Liemert-Kienle12}
Liemert, A. and Kienle, A.,
``Infinite space Green's function of the time-dependent radiative transfer equation,''
{\it Biomed. Opt. Exp.} {\bf 3}, 543--551 (2012).

\bibitem{Lin-Xu07}
Lin, Y. and Xu, C.,
``Finite difference/spectral approximations for the time-fractional diffusion equation,''
{\it J. Comp. Phys.} {\bf 225}, 1533--1552 (2007).

\bibitem{Luchko09}
Luchko, Y.,
``Maximum principle for the generalized time-fractional diffusion equation,''
{\it J. Math. Anal. Appl.} {\bf 351}, 218--223 (2009).

\bibitem{Luchko10}
Luchko, Y.,
``Some uniqueness and existence results for the initial-boundary-value problems for the generalized time-fractional diffusion equation,''
{\it Comp. Math. Appl.} {\bf 59}, 1766--1772 (2010).

\bibitem{Luchko11}
Luchko, Y.,
``Initial-boundary-value problems for the generalized multi-term time-fractional diffusion equation,''
{\it J. Math. Anal. Appl.} {\bf 374}, 538--548 (2011).

\bibitem{Luchko12}
Luchko, Y.,
``Initial-boundary-value problems for the one-dimensional time-fractional diffusion equation,''
{\it Fract. Cal. Appl. Anal.} {\bf 15}, 141--160 (2012).

\bibitem{Mainardi96}
Mainardi, F.,
``The fundamental solutions for the fractional diffusion-wave equation,''
{\it Appl. Math. Lett.} {\bf 9}, 23--28 (1996).

\bibitem{Mainardi-Luchko-Pagnini01}
Mainardi, F., Luchko, Y., Pagnini, G.,
``The fundamental solution of the space-time fractional diffusion equation,''
{\it Fract. Cal. Appl. Anal.} {\bf 4}, 153--192 (2001).

\bibitem{Mellet10}
Mellet, A.,
``Fractional diffusion limit for collisional kinetic equations: A moments method,''
{\it Indiana Univ. Math. J.} {\bf 59}, 1333--1360 (2010).

\bibitem{Mellet-Mischler-Mouhot11}
Mellet, A., Mischler, S., and Mouhot, C.,
``Fractional diffusion limit for collisional kinetic equations,''
{\it Arch. Rational Mech. Anal.} {\bf 199}, 493--525 (2011).

\bibitem{Metzler-Klafter00}
Metzler, R. and Klafter, J.,
``The random walk's guide to anomalous diffusion: a fractional dynamics approach,''
{\it Phys. Rep.} {\bf 339}, 1--77 (2000).

\bibitem{Metzler-Klafter04}
Metzler, R. and Klafter, J.,
``The restaurant at the end of the random walk: recent developments in the description of anomalous transport by fractional dynamics,''
{\it J. Phys. A: Math. Gen.} {\bf 37}, R161--R208 (2004).

\bibitem{Metzler-etal14}
Metzler, R., Jeon, J.-H., Cherstvya, A. G., and Barkaid, E.,
``Anomalous diffusion models and their properties: non-stationarity, non-ergodicity, and ageing at the centenary of single particle tracking,''
{\it Phys. Chem. Chem. Phys.} {\bf 16}, 24128--24164 (2014).

\bibitem{Panasyuk-etal06}
Panasyuk, G., Schotland, J. C., and Markel, V. A.,
``Radiative transport equation in rotated reference frames,''
{\it J. Phys. A: Math. Gen.} {\bf 39}, 115--137 (2006).

\bibitem{Podlubny99}
Podlubny, I.,
{\it Fractional Differential Equations}
(Academic Press, 1999).

\bibitem{Ryzhik-etal96}
Ryzhik, L., Papanicolaou, G., and Keller, J. B.,
``Transport equations for elastic and other waves in random media,''
{\it Wave Motion} {\bf 24}, 327--370 (1996).

\bibitem{Sakamoto-Yamamoto11}
Sakamoto, K. and Yamamoto, M.,
``Initial value/boundary value problems for fractional diffusion-wave equations and applications to some inverse problems,''
{\it J. Math. Anal. Appl.} {\bf 382}, 426--447 (2011).

\bibitem{Samko93}
Samko, S. G., Kilbas, A. A., and Marichev, O. I.,
{\it Fractional integrals and derivatives: theory and applications} 
(Gordon and Breach Science, 1993).

\bibitem{Sokolov-Klafter-Blumen02}
Sokolov, I., Klafter, J., and Blumen, A.,
``Fractional Kinetics,''
{\it Physics Today} {\bf 55}, 48--54 (2002).

\bibitem{Williams92}
Williams, M. M. R.,
``Stochastic problems in the transport of radioactive nuclides in fractured rock,''
{\it Nucl. Sci. Eng.} {\bf 112}, 215--230 (1992).

\bibitem{Williams93}
Williams, M. M. R.,
``Radionuclide transport in fractured rock a new model: application and discussion,''
{\it Ann. Nucl. Energy} {\bf 20}, 279--297 (1993).

\end{thebibliography}


\end{document}